\documentclass[aps,prl,a4paper,twocolumn,showpacs,floatfix,citeautoscript]{revtex4}
\usepackage{graphicx}
\usepackage{amsfonts}
\usepackage{amsmath}
\usepackage{amssymb}
\usepackage{bm}
\usepackage{dcolumn}
\usepackage{hyperref}
\usepackage[usenames]{color}
\hypersetup{pdfborder=0 0 0,colorlinks=true,citecolor=blue,linkcolor=blue}
\begin{document}

\title{Theory of spin-orbit coupling at LaAlO$_3$/SrTiO$_3$ interfaces and SrTiO$_3$ surfaces}
\author{Zhicheng Zhong, Anna T{\'o}th, and Karsten Held}
\affiliation{Institute of Solid State Physics, Vienna University of Technology, A-1040 Vienna, Austria}
\date{\today}

\begin{abstract}
The theoretical understanding of the spin-orbit coupling (SOC) effects at LaAlO$_{3}$/SrTiO$_{3}$ interfaces and SrTiO$_{3}$ surfaces is still in its infancy. We perform first-principles density-functional-theory calculations and derive from these a simple tight-binding Hamiltonian, through a Wannier function projection and group theoretical analysis. We find striking differences to the standard Rashba theory for spin-orbit
coupling in semiconductor heterostructures due to multi-orbital effects: by far the biggest SOC effect is at the crossing point of the $xy$ and $yz$ (or $zx$) orbitals; and around the $\Gamma$ point a Rashba spin splitting with a cubic dependence on the wave vector $\vec{k}$ is possible.
\end{abstract}
\maketitle
\textcolor{red}{Introduction:} Very recently, Caviglia {\it et al.}\cite{Caviglia:prl10b} and Ben Shalom {\it et al.}\cite{Shalom:prl10b} studied the magnetotransport properties of the high mobility two-dimensional electron gas (2DEG) at the interface between two insulating perovskite oxides LaAlO$_{3}$ (LAO) and SrTiO$_{3}$ (STO)\cite{Ohtomo:nat04}. They found strong spin-orbit coupling (SOC) effects, whose magnitudes can even be tuned by gate voltages. This strong SOC has been the basis for many subsequent theoretical and experimental studies\cite{Michaeli:prl11, Reyren:prl12,Fidkowski:arxiv12,Caprara:arxiv12,Fischer:arxiv12,Fete:arxiv12}. Despite its increasing importance, a clear physical picture of the SOC effects in LAO/STO interfaces is still missing.

SOC generally originates from the relativistic correction $(\hbar / 2m^{2}_{e}c^{2})(\nabla V \times \vec{p})\cdot \vec{s}$ to the Schr\"odinger equation, with $m_{e}$ being the free electron mass and $V$ the potential in which the electrons move with momentum $ \vec{p}$ and spin $\vec{s}$. If $V(r)$ has spherical symmetry like in atoms or approximately in solids, the form can be reduced to $\xi(r)\vec{l}\cdot \vec{s}$, where $\xi(r)$ denotes the strength of the atomic SOC, and $\vec{l}$ and $\vec{s}$ are the orbital and spin angular momenta of the electron. Thus SOC lifts orbital and spin degeneracies. For a cubic perovskite such as STO, the six initially degenerate $t_{2g}$ orbitals at the $\Gamma$ point are indeed split by SOC\cite{Mattheiss:prb72, Janotti:prb11, Bistritzer:prb11,Marel:prb11}, but Kramers degeneracy (time reversal symmetry) is preserved. Because bulk STO has both crystal inversion symmetry $\varepsilon(\vec{k}, \uparrow)=\varepsilon(-\vec{k}, \uparrow)$ and time reversal symmetry $\varepsilon(\vec{k}, \uparrow)=\varepsilon(-\vec{k}, \downarrow)$, the energy $\varepsilon$ at wave vector $\vec{k}$ is still spin-degenerate.

This changes for a LAO/STO interface since the inversion symmetry is broken so that the SOC lifts the spin degeneracy of the 2DEG. This two dimensional SOC effect is known as Rashba spin splitting, which has been widely studied in semiconductor heterostructures\cite{Winkler:book, Nitta:prl97} and metal surfaces\cite{LaShell:prl96}. Assuming a nearly free 2DEG with effective mass $m$, its two spin components will be split by $\Delta_{R} = 2\alpha_{R} k$, where the Rashba coefficient $\alpha_{R}= (\hbar / 4m^{2}c^{2}){\rm d}V(z)/{\rm d}z$ depends on the potential gradient in the $z$ direction (perpendicular to the interface). Naively, since an electric field gives rise to an electrostatic potential gradient, $\Delta_{R}$ seems to be simply proportional to the gate voltage. However, a typical electric field in experiments is $\sim$100 V/mm yielding according to the formula above $\Delta_{R}\sim 10^{-8}$meV\cite{Winkler:book}, which is much smaller than the measured values of the order of meV\cite{Caviglia:prl10b, Shalom:prl10b}. This is because the assumption of nearly free 2DEG, which ignores the region of ion cores and the asymmetric feature of the interface wavefunction, is too simple. In reality, the expression for $\Delta_{R}$ is much more complicated \cite{Lommer:prl88,Schapers:jap98,Nicolay:prb01,Bihlmayer20063888,Nagano:jpcm09} and usually treated as a fitting parameter in semiconductor heterostructures and metal surfaces.

The aim of this Letter is the theoretical description of SOC in oxide heterostructures and surfaces. From density-functional-theory (DFT) calculations we derive a tight-binding (TB) Hamiltonian for the low energy $t_{2g}$-orbitals and their spin-splitting. In particular, we show that besides the standard $k$-linear Rashba spin splitting, there can be a $k$-cubic spin splitting around the $\Gamma$ point due to multi-orbital effects. A much larger spin splitting occurs at the crossing point of the $xy$ and $yz$ (or $zx$) orbitals.

\textcolor{red}{Method:}
Besides bulk STO, we calculate 
(i) LAO/STO (1.5/6.5 layers) \cite{Popovic:prl08,Delugas:prl11} which is symmetric with two $n$-type interfaces,
(ii) LAO/STO (4/4) and (1/1) which is asymmetric with $n$- and $p$-type interfaces \cite{zhong:prb10},
(iii) vacuum/LAO/STO (3/1/4) which has a single $n$-type interface\cite{Pentcheva:prl09,Son:prb09},
(iv) vacuum/STO (3/7.5) with a SrO terminated surface.
We fix the in-plane lattice constant of the supercells to the calculated equilibrium value of STO, and optimize the internal coordinates. The DFT calculations have been done using the WIEN2k code with the generalized gradient approximation (GGA)\cite{WIEN2k,PerdewPRL96}. The SOC is included as a perturbation using the scalar-relativistic eigenfunctions of the valence states. Through a projection onto maximally localized Wannier orbitals\cite{Mostofi2008685,Kune20101888,zhong:epl12}, we construct a realistic TB model, and in particular, we develop a way to describe the interface asymmetry.
\begin{figure}[t!]
\includegraphics[width=\columnwidth]{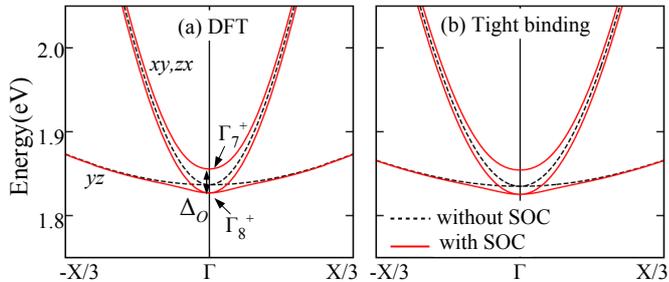}
\caption{Band structure of $t_{2g}$ orbitals in bulk SrTiO$_3$ calculated by (a) DFT and by (b) a TB model derived in this paper. In the absence of spin-orbit coupling, $yz$, $zx$ and $xy$ are degenerate at the $\Gamma$ point. SOC splits the six-fold degenerate orbitals into $\Gamma_{7}^{+}$ and $\Gamma_{8}^{+}$ states separated by $\Delta_{O}=$29~meV.}
\label{Fig1}
\end{figure}

\textcolor{red}{DFT results for STO:} Bulk SrTiO$_3$ is a band insulator with an energy gap between occupied O$_{2p}$ bands and unoccupied Ti$_{3d}$ $t_{2g}$ ($yz, zx, xy$) bands. The $t_{2g}$ band structure calculated by DFT is shown in Fig.\ref{Fig1}(a). In the absence of SOC, the three $t_{2g}$ orbitals are degenerate at $\Gamma (0, 0, 0)$ due to an octahedral $O_h$ crystal field around the Ti atoms in a perfect perovskite structure. Around $\Gamma$ the bands can be fit by a parabolic function of the form $\hbar ^{2} k^{2}/2m^*$ where the effective mass $m^*$ depends on the orbital and the direction.

Along the $\Gamma-$X$(\pi, 0, 0)$ direction (here in units of $1/a$ with $a$=3.92\AA\ being the calculated lattice constant of STO), the $yz$ band has a small energy dispersion corresponding to a heavy mass of $6.8m_{e}$. In contrast, the $zx$ and $xy$ orbitals have the same, large energy dispersion and a light effective mass of $0.41m_{e}$. Including the SOC, the six-fold degenerate orbitals are split into a doubly and a four-fold degenerate level with an orbital splitting of $\Delta_{O}=$ 29~meV \cite{Marel:prb11}, see Fig.\ref{Fig1}. The energy dispersion of the resulting orbitals is now considerably different from the initial orbitals. Consequently, the corresponding effective masses around $\Gamma$ are changed to $1.39$, $0.41$ and $0.53m_{e}$, respectively.

\textcolor{red}{TB Hamiltonian for STO:} To understand the DFT results, we use a Wannier projection to obtain the local energy and hopping terms of the $t_{2g}$ orbitals.
 Without SOC, the constructed  three-band TB Hamiltonian of bulk STO $H_{0}^{b}$ can be expressed in the $t_{2g}$ basis in a matrix form: one of the diagonal terms is $\langle xy|H_{0}^{b}|xy\rangle=\varepsilon_{0}^{xy}-2t_{1}cosk_{x}-2t_{1}cosk_{y}-2t_{2}cosk_{z}-4t_{3}cosk_{x}cosk_{y}$; the other two follow  by exchanging the $x,y,z$ indices.
The local energy term is $\varepsilon_{0}=3.31$eV for all three orbitals. The large hopping $t_{1}=0.277$eV stems from the large $xy$ intra-orbital hopping integral along the $x$ and $y$ direction; it is due to two lobes of $xy$ orbitals at the nearest neighbor sites pointing to each other along the two directions. In contrast, $t_{2}=0.031$eV and $t_{3}=0.076$eV indicate a much smaller hopping integral along the $z$ and $(1, 1, 0)$ direction respectively. We find all orbital-off-diagonal (inter-orbital hopping) terms to be negligible. The TB energy dispersion for the three orbitals is plotted in Fig.\ref{Fig1}(b), showing a good agreement with the DFT results. 

We include SOC at the atomic level as $H_{\xi}=\xi \vec{l}\cdot \vec{s}$, where $\xi$ is the atomic SOC strength and depends on atomic numbers. Under the $O_h$ crystal field, the six spinful $t_{2g}$ orbitals break into a doublet $|\Gamma_7^+, \pm\frac12\rangle \equiv \frac{1}{\sqrt{3}}\left(-i\,yz|\downarrow,\uparrow\rangle \pm zx|\downarrow,\uparrow\rangle \mp i\,xy|\uparrow,\downarrow\rangle\right)$, and a quartet $|\Gamma_8^+, \pm \frac12\rangle  \equiv  \frac{1}{\sqrt{2}}\left(\mp i\,yz-zx\right)|\downarrow,\uparrow\rangle$, $|\Gamma_8^+,\pm \frac32\rangle  \equiv \frac{1}{\sqrt{6}}\left(\pm i\,yz|\uparrow,\downarrow\rangle+zx|\uparrow,\downarrow\rangle+2i\,xy|\downarrow,\uparrow\rangle\right)$\cite{Koster}. 

$H_\xi$ lifts their degeneracy at the $\Gamma$ point and is diagonal in the $\Gamma_7^+, \Gamma_8^+$ basis with eigenvalues of $\xi$ and $-\xi$/2, respectively. We set $\xi=2\Delta_{O} /3=19.3$~meV, which leads to the same orbital splitting as the DFT results. In the original $t_{2g}$ basis ($yz|\!\uparrow\rangle$, $yz|\!\downarrow\rangle$, $zx|\!\uparrow\rangle$, $zx|\!\downarrow\rangle$, $xy|\!\uparrow\rangle$, $xy|\!\downarrow\rangle$), $H_{\xi}$ has off-diagonal terms and reads
\begin{equation*}
 \frac{\xi}{2}\left(\begin{array}{cccccc}
0 & 0 & i & 0 & 0 & -1\\
0 & 0 & 0 & -i & 1 & 0\\
-i & 0 & 0 & 0 & 0 & i\\
0 & i & 0 & 0 & i & 0\\
0 & 1 & 0 & -i & 0 & 0\\
-1 & 0 & -i & 0 & 0 & 0\end{array}\right).
\end{equation*}
$H_{0}^{b}+H_{\xi}$ is the TB Hamiltonian of bulk STO including SOC. Its bandstructure is shown as a solid line in Fig.\ref{Fig1}(b) and agrees well with the DFT results. This model allows for a deeper understanding of the SOC effects:  the SOC eigenstates are admixtures of the $yz$, $zx$ and $xy$ orbitals, which explains the significant changes of the effective masses.

\begin{figure}[t!]
\includegraphics[width=\columnwidth]{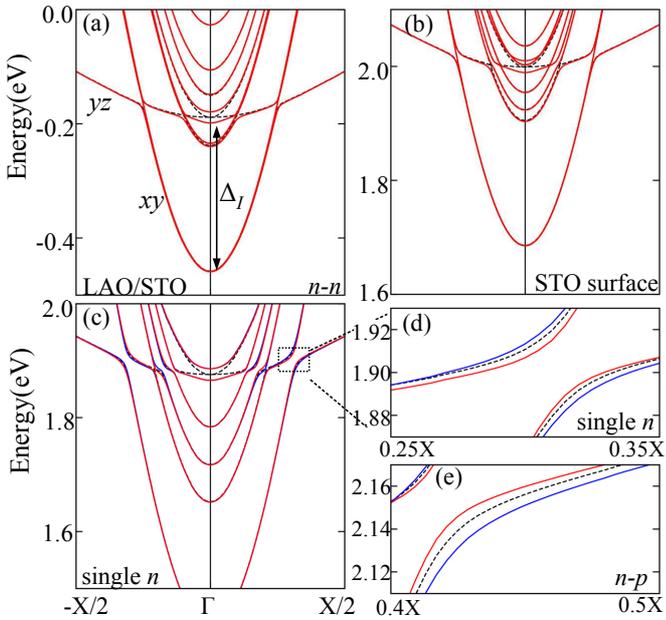}
\caption{DFT calculated band structure of (a) LAO/STO with two symmetric $n$-type interfaces, (b) SrO terminated STO surface, and (c) vacuum/LAO/STO with a single $n$-type interface.
The region with biggest spin splitting around the $xy$-$yz$ crossing point is magnified in (d) and for a LAO/STO (4/4) with asymmetric $n$-$p$ interfaces in (e). $\Delta_{I}$ denotes the interface induced orbital splitting energy; the black dashed line is without SOC; the red and blue solid lines are the two opposite spin channels with SOC.}
\label{Fig2}
\end{figure}

\textcolor{red}{DFT results for interfaces:} The band structure of LAO/STO (1.5/6.5) calculated by DFT is shown in Fig.\ref{Fig2}(a). Without SOC, similar to bulk STO, all bands exhibit a parabolic-like behavior. In the $x$ direction, the $yz$ band is the flattest (heaviest); at $\Gamma$, it is degenerate with the $zx$. Due to the interface, the $xy$ band is $\Delta_{I}=250$~meV lower in energy  at $\Gamma$ than the degenerate $yz$ and $zx$ bands. The splitting $\Delta_{I}$ is the most notable feature of the heterostructure\cite{Popovic:prl08, Delugas:prl11,Son:prb09}. It is not mainly a crystal field effect, but originates from the vanishing of the hopping from the interface Ti $yz$ to LAO along the $z$ direction \cite{zhong:epl12}. Consistently, a similar behavior is expected in a SrO terminated STO surface, and indeed we obtain it with $\Delta_{I}=320$~meV, see Fig.\ref{Fig2}(b). Our calculated splittings are qualitatively consistent with the ARPES measurements of STO surfaces\cite{Santander:nat11, Meevasana:natm11}.

\begin{figure}[t!]
\includegraphics[width=\columnwidth]{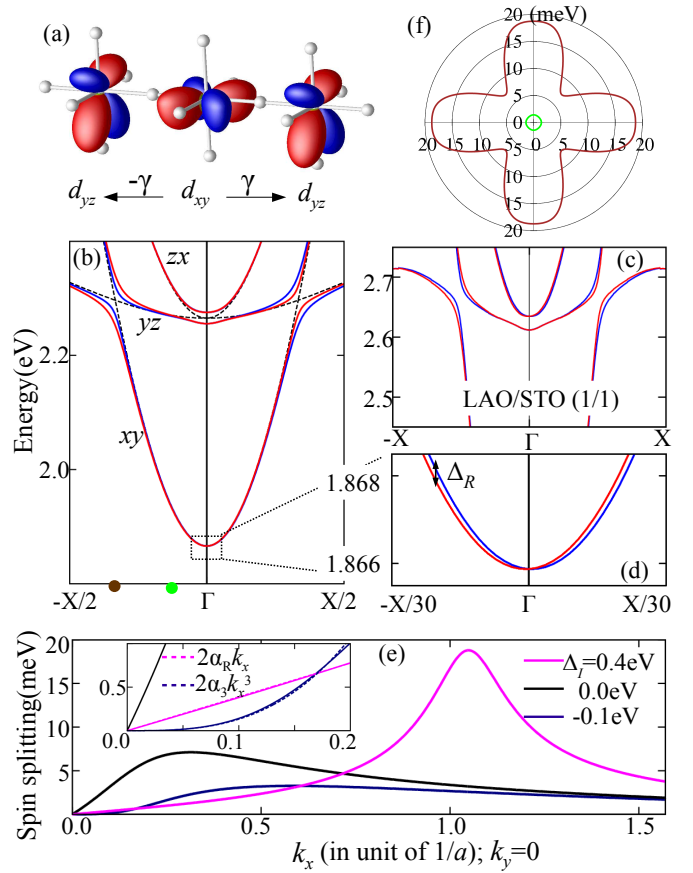}
\caption{(a) Schematic figure of the orbital deformation at the interface resulting in an anti-symmetric hopping term $\gamma$. Band structures calculated by (b) TB Hamiltonian $H^{i}_{0}+H_{\xi}+H_{\gamma}$, and by (c) DFT for comparison; (d) the standard Rashba spin splitting $\Delta_{R}$ can be observed after zooming in the band structure of panel (b); red and blue indicate opposite spin. (e) Spin-splitting in the lowest band of the TB Hamiltonian for different $\Delta_I$. The inset shows a $k_{x}$-linear (-cubic) spin splitting for positive (negative) $\Delta_I$. (f) Angular dependence of the spin splitting at the standard Rashba region (green) and the $xy$-$yz$/$zx$ crossing region (brown) at the $\vec{k}$ points displayed in (b).}
\label{Fig3}
\end{figure}

Including SOC does not influence the $xy$ band very much. It splits the degenerate $yz$ and $zx$ orbitals with $\Delta_{O}=19$~meV at $\Gamma$, see Fig.\ref{Fig2}(a). The {\em ab initio} calculated $\Delta_{O}$ is qualitatively consistent with experiment, albeit smaller than its experimental value \cite{Santander:nat11} $\Delta_{O}=60$~meV. Around $\Gamma$, the effective masses for $xy$ and the resulting two states are 0.48, 1.14 and 0.72 $m_{e}$, respectively. For the asymmetric case Fig.\ref{Fig2}(c) the SOC also results in a spin splitting which is most noticeable at the $xy$-$yz$ crossing region where it is up to 18 meV, see Fig.\ref{Fig2}(d,e). This spin splitting is a multi-orbital effect, very different from the standard Rashba spin splitting of single orbital. For a better understanding, we now construct a TB Hamiltonian.

\textcolor{red}{Spin splitting at the interface layer:} Without SOC, a model Hamiltonian $H_{0}^{i}$ can describe the interface hopping and the induced splitting $\Delta_{I}$. In contrast to $H_{0}^{b}$, the hopping terms of $H_{0}^{i}$ in direction $z$ essentially vanish. The diagonal term for $xy$ is hence $\varepsilon^{xy}-2t_{1}\cos k_{x}-2t_{1}\cos k_{y}-t_{2}-4t_{3}\cos k_{x}\cos k_{y}$, while that for the $yz$ (and $zx$) orbital is $\varepsilon^{yz}-2t_{2}\cos k_{x}-2t_{1}\cos k_{y}-t_{1}-2t_{3}\cos k_{y}$. The local energy terms $\varepsilon^{xy/yz}$ will be influenced by the interface crystal field, electron filling and confinement\cite{Delugas:prl11,Son:prb09,zhong:epl12}. For simplicity, we approximate these by the bulk value $\varepsilon_{0}$. Thus, $\Delta_{I}= t_{1}-t_{2}+2t_{3}=$0.4eV which is comparable to the DFT results. At the interface the $O_h$ symmetry breaks down to $C_{4v}$, and we can use the same atomic SOC $H_{\xi}$ matrix as before, since under $C_{4v}$ the $\Gamma_7^+$ doublet does not break whereas the $\Gamma_8^+$ quartet breaks into $\Gamma_6^+\oplus\Gamma_7^+$, with the same set of basis functions as given previously. 
The $H_{0}^{i}+H_{\xi}$ Hamiltonian gives an atomic SOC induced orbital splitting about $\Delta_O=\sqrt{5} \xi/{2}$ at $\Gamma$ similar to Figs.\ \ref{Fig2} (a-c).
However  $H_{0}^{i}+H_{\xi}$ does not contain any terms breaking the interface inversion symmetry, and hence it does not include the Rashba spin splitting.

To this end, we introduce a term $H_{\gamma}$ to describe the broken inversion symmetry at the interface, a key component for Rashba spin splitting. The essential physics of this term was analyzed by Lashell {\it et al.} \cite{LaShell:prl96} and then introduced by Petersen {\it et al.}\cite{Petersen:ss00} to construct a TB model for the Rashba effects of $s$-$p$ orbitals in metal surfaces. To our knowledge, there and in other publications, $H_{\gamma}$ was always treated as a parameter and hence its utility and importance are strongly limited. In this study, we project the DFT results above
onto maximally localized Wannier orbitals \cite{zhong:epl12} and then directly extract the spin independent hopping term $H_{\gamma}$
\begin{equation*} 
\gamma\left(\begin{array}{cccccc}
0   & 0  & 2i\sin k_{x}\\
0  & 0  & 2i\sin k_{y}\\
-2i\sin k_{x}  & -2i\sin k_{y}  & 0\end{array}\right)
\end{equation*}
describing inter-orbital hopping terms due to the interface asymmetry. The key hopping term is $\gamma=\langle xy|H|yz(R)\rangle$, where $R$ is the nearest neighbor in $x$ direction. As shown in the schematic Fig.\ref{Fig3}(a), $\gamma$ is an anti-symmetric hopping between $xy$ and $yz$ orbitals along the $x$ direction. Its origin is the interface asymmetry deforming the orbital lobes of the interface layer. We find $\gamma \sim 20$meV at the $n$-type interface for all geometries, and hence take this value in the model. Let us note $\gamma$ drops quickly in  the second and further layers towards its bulk value  $\gamma=0$.

The combined model Hamiltonian $H_{0}^{i}+H_{\xi}+H_{\gamma}$, including Rashba effects, is expressed in the $t_{2g}$ basis by a $6\times6$ matrix, where $H_{0}^{i}$ describes the interface hopping and splitting $\Delta_{I}$, $H_{\xi}$ includes the atomic SOC and accounts for the orbital splitting of $\Delta_{O}$, and $H_{\gamma}$ describes the interface asymmetry. The first effect is a standard Rashba-type of spin splitting in the single $xy$ band. It splits a single parabola around the minimum at $\Gamma$ into two parabolas with opposite spin, see 
 Fig.\ref{Fig3}(d). By downfolding the matrix onto an effective Hamiltonian for the $xy$ band, we obtain an analytical expression for the spin splitting $\Delta_R = 2 \alpha_R k_{x}$ with 
$\alpha_R = 2a\xi \gamma/\Delta_{I}=0.76\times10^{-2}$eV\AA\ for $\Delta_{I}=0.4$eV, $\xi=19.3$meV, $\gamma=20$meV. Note that $\Delta_{I}$ depends strongly on the details of the interface and hence
 $\alpha_R$ can be up to 8 times larger at $\Delta_{I}=0$ where the formula above does not hold anymore, see Fig.\ref{Fig3}(e).
This well agrees with the experimental magnitude of the Rashba spin splitting $\alpha_R = 1-5\times10^{-2}$eV\AA\cite{Caviglia:prl10b}.

If we turn $\Delta_{I}$ negative, which is possible by interface engineering
\cite{zhong:epl12}, the lowest band is a mixture of $yz$ and $zx$. In this situation
there is no standard $k$-linear Rashba effect any more but we obtain
a spin splitting  $2 \alpha_{3} k_{x}^3$ with $\alpha_{3}= 4$eV\AA$^3$,
see Fig.\ref{Fig3}(e). Hence, the TB model also explains qualitatively and quantitatively 
the unusual $k$-cubic spin splitting reported in \cite{Nakamura:prl12} in a single framework,
reconciling the puzzling discrepancy between experiments \cite{Nakamura:prl12} and \cite{Caviglia:prl10b}.

An even much bigger spin splitting 18meV occurs however at the $xy$-$yz$ crossing  point, see Figs.\ref{Fig3}(b,c,e) and Figs.\ref{Fig2}(c-e).
Unlike the isotropic splitting  $\Delta_{R}$ around $\Gamma$, this spin splitting is not only much larger but also anisotropic, see Fig.\ref{Fig3}(f). This multiband effect is a particularity of transition metal oxides and not occurring in semiconductors or metal surfaces. Experimentally,  a similar anisotropy has been observed in LAO/STO heterostructures with a particular strong SOC effect \cite{Flekser:prb12}.

An important aspects of our study is also that the external electric field \cite{Caviglia:prl10b,Shalom:prl10b} does not significantly  tune the SOC directly.
As mentioned in the introduction, its direct contribution is too small. Even  
without it, we obtain the correct magnitude of the spin splitting.
Nonetheless, the spin splitting depends on the electric field \cite{Caviglia:prl10b,Shalom:prl10b}.
The explanation for this is an indirect effect: the electric field tunes the carrier densities \cite{Thiel:sc06, Caviglia:nat08,Bell:prl09,Shalom:prl10,Caviglia:prl10c}, band filling \cite{Copie:prl09,Joshua:arxiv11, Delugas:prl11,Son:prb09}, and the effective $\gamma$.  The multi-orbital complexity might account for the discrepancy of the two reported spin energies tuned by gate voltages \cite{Caviglia:prl10b,Shalom:prl10b}.

\textcolor{red}{Conclusion:} We performed first principle calculations and developed a realistic three-band ($xy$, $yz$ and $zx$) model for SOC effects at  LAO/STO interfaces and STO surfaces. The key ingredients to the spin splitting are the atomic SOC and the interface asymmetry, which enters via asymmetric $t_{2g}$ orbital lobes. The $xy$ orbital around $\Gamma$ exhibits the standard Rashba spin splitting $2 \alpha_R k$ with $\alpha_R = 2a\xi \gamma/\Delta_{I}\sim 10^{-2}$ eV\AA; in contrast, for negative $\Delta_{I}$
there is instead a $k$-cubic dependence spin splitting in the lowest band around $\Gamma$. As $\Delta_{I}$ depends on the particular surface or interface, this
solves the experimental controversy regarding linear or cubic Rashba splitting.
Even more importantly, we find an unusually large spin splitting 18meV at the crossing point of $xy$ and $yz/zx$ orbitals. Our results indicate that  LAO/STO has peculiar SOC properties arising from the multi-orbitals character  which are absent in the standard single-band description as for the nearly free 2DEG in semiconductor heterostructures.

We are grateful to G. Sangiovanni, P. Wissgott, R. Arita, V. I. Anisimov, and P. J. Kelly for useful discussions. Z.Z.\ and K.H.\ acknowledge funding from the SFB ViCoM (Austrian Science Fund project ID F4103-N13),
 A.T.\ from the European Research Council under
Grant No.\ FP7-ERC-227378 and from the EU-Indian network MONAMI and the Austrian Ministry for Science and Research (BM.W\_F).
 Calculations have been done on the Vienna Scientific Cluster~(VSC).

\end{document}